\documentclass[]{aa}
\usepackage{graphicx}
\usepackage[round]{natbib}
\usepackage{epsfig}
\usepackage{aalongtable}

\begin{document}
\title{New method for determining the thickness of non-edge-on disk galaxies}
\author{Yinghe Zhao, Qiuhe Peng \and TaoHu}

\offprints{Yinghe Zhao} \institute{Department of Astronomy,
Nanjing University, Nanjing 210093, China\\
     \email{yhzhao, qhpeng, taohu@nju.edu.cn}}

\abstract{ In this paper we report a new method for determining
the thickness of the non-edge-on disk galaxies. Our method is
based on the observation comparing with the theoretical researches
of the distribution of the vertical velocity dispersion, which is
obtained from the solutions of three dimensional Poisson's
equation and the galactic dynamical equation. This method also
allows us to investigate the mass-to-light ratio of the disk. As
examples, the thickness and mass-to-light ratio of two disk
galaxies, NGC 1566 and NGC 5247, which have been extensively
studied in spectroscopy, have been calculated and the results are
presented.

}

\authorrunning{Zhao et al.}
\titlerunning{New method for determining the thickness of disk galaxies}

\maketitle

\date{Received~~~~~~ ; accepted~~~~~~~ }

\section{Introduction}

The thickness of disk galaxies is a very important parameter for
understanding these objects, although all of which are very thin
comparing with the length of the disk. Sancisi and Allen (1979)
estimated the thickness of the edge-on Sb galaxy NGC 891, on the
basis of observation of neutral hydrogen. Van der Kruit \& Searle
(1981a) proposed a model for light distribution in the disks of
edge-on spiral galaxies, assuming that a galaxy has a locally
isothermal, self-gravitating, truncated and exponential disk. The
model has the feature of being isothermal in $z$-direction at all
radii with a scale parameter of $z_0$ and has an exponential
dependence of surface brightness on $r$ with scale length of
$r_d$. The space luminosity of this model can be described by
\begin{equation}
L(r,z) = L_0 e^{ - r/r_d } \sec h^2 (z/z_0 )
\end{equation}
With this model, van der Kruit \& Searle (1981a, 1981b, 1982a,
1982b, named KS hereafter) determined $r_d$ and $z_0$ for several
edge-on galaxies without an appreciable bulge. Unfortunately, this
method may not be suitable for general disk galaxies, e.g. face-on
galaxies.

Peng (1988) put forward a method to measure the thickness of
face-on galaxies on the basis of asymptotic formula of the
disturbed gravitational potential. And a revised method based on
the exact integral expression for disturbed gravitational
potential has been presented by Zhao et al. (2004) to estimate the
thickness of face-on disk galaxies. But the results obtained by
these two correlative methods should be checked with the results
gotten by other independent ways.

In this paper, we present the solution of the Jeans equation along
$z$-direction first and then describe how to use this solution to
determine the thickness of face-on disk galaxies. In Section 3, we
show the applications of our method to two near face-on galaxies,
NGC 1566 and NGC 5247, which have been extensively studied in
spectroscopy. The results are also presented. And a brief
discussion is given in the last section of this paper.

\section{Method}
\subsection{The distribution of the vertical velocity dispersion}
For a stable, axisymmetric galaxy, the Jeans equation along the
$z$-direction is
\begin{equation}
\frac{\partial }{{\partial z}}(\rho \left\langle {V_z^2 }
\right\rangle ) =  - \rho \frac{{\partial \varphi }}{{\partial
z}},
\end{equation}
where $\left\langle {V_z^2 } \right\rangle$ is square of the
$z$-direction velocity dispersion.

As the first of the fundamental assumptions, we accept Parenago's
density distribution law along the $z$-direction for a finite
thickness galaxy,
\begin{equation}
\rho (r,z) = \rho (r,0)e^{ - \alpha |z|}  = \frac{\alpha
}{2}\sigma (r)e^{ - \alpha |z|},
\end{equation}
where $\alpha$ is a parameter the reciprocal of which is half of
the effective thickness, $H$, of a galaxy. $\alpha$ may be taken
as a constant, for the scale height is basically independent of
the radius at least for the thick disk of our Galaxy (e.g. see
Freeman, 1987). $\sigma(r)$ is its surface density. de Grijs \&
van der Kruit (1996) have also confirmed that the simple
exponential fit of the vertical stellar light distribution is a
good approximation. Hence, the potential caused by this density
can be found by solving the Poisson's equation of the potential
\begin{equation}
\nabla ^2 \varphi (r,z) = 2\pi G\alpha \sigma (r)e^{ - \alpha
|z|}.
\end{equation}

The rigorous solution of the potential has been given by Peng et
al. (1978)
 \begin{equation}
 \begin{array}{l}
 \varphi (r,z) =  - \pi G\alpha  \\
\ \ \ \ \ \ \ \ \ \ \ \
  \cdot \int_0^\infty  {\frac{2}{{\beta ^2  - \alpha ^2 }}(\beta e^{ - \alpha |z|}  - \alpha e^{ - \beta |z|} )J_0 (\beta r)} s(\beta )d\beta  \\
 \end{array}
\end{equation}
where
\begin{equation}
 s(\beta ) = \int_0^\infty  r J_0 (\beta
r)\sigma (r)dr.
\end{equation}
And $J_0(\beta r)$ is the well known Bessel function of order 0.

Then using equation (5), we integrate equation (2) from $z$ to
$\infty$ ($z \geq 0$) at both sides of the equal sign
respectively, it follows (Huang, Huang \& Peng 1979)
\begin{equation}
\begin{array}{l}
 (\rho \left\langle {V_z^2 } \right\rangle )_{z \ge 0}  =  - 2\pi G\alpha ^2 \rho (r,0) \\
\\
  \cdot \int_0^\infty  {\frac{\beta }{{\beta ^2  - \alpha ^2 }}} \left[ {\frac{1}{{\alpha
  + \beta }}e^{ - (\alpha  + \beta )z}  - \frac{1}{{2\alpha }}e^{ - 2\alpha z} } \right]J_0 (\beta r)s(\beta )d\beta  \\
 \end{array}
\end{equation}
Hence, on the galactic plane ($z=0$ ), $\left\langle {V_z^2 }
\right\rangle$ could be written as
\begin{equation}
\begin{array}{c}
 \left\langle {V_z^2 } \right\rangle _{(z = 0)}  = \pi G\alpha \int_0^\infty  {\frac{\beta }{{(\alpha  + \beta )^2 }}} J_0 (\beta r)s(\beta )d\beta  \\
\\
  = \frac{{\pi G}}{\alpha }\sigma (r)[1 - P(r)] \\
 \end{array}
\end{equation}
where
\begin{equation}
\begin{array}{l}
P(r,\alpha) = \frac{1}{{\sigma (r)}}\int_0^\infty  {\beta \zeta
(\alpha ,\beta )J_0 (\beta r)s(\beta )d\beta },\\
\\
{\rm{ }}\zeta (\alpha ,\beta ) = 1 - \frac{1}{{(1 + \beta /\alpha
)^2 }}.
\end{array}
\end{equation}

\subsection{The thickness parameter $\alpha$}
By assuming that a galaxy has an infinitesimally thin disk,
Freeman (1970) studied 36 spiral and S0 galaxies with surface
photometry and showed that the radial light distribution in the
disks of spiral galaxies can be described by an exponentially
decreasing surface-brightness with increasing galactocentric
radius. Along the vertical direction, de Grijs \& van der Kruit
(1996) have shown that a simple exponential fits turned out to be
good approximations of the stellar light distribution. Thus, the
model to account for the space-luminosity can be described by
\begin{equation}
L(r,z) = L_0 e^{ - r/r_d } e^{ - \alpha |z|},
\end{equation}
where $r$ is the position along the major axis, $L_0$ the central
space luminosity in the plane of the galaxy, $r_d$ the scale
length of the disk. From this we calculate the face-on
distribution of surface-brightness
\begin{equation}
I (r) = I _0 e^{ - r/r_d },
\end{equation}
where $I_0$ is the central surface brightness of the disk.

Making use of the observed constant color index (Van der Kruit \&
Searle 1981b, 1982a), Van der Kruit \& Freeman (1986) pointed out
that the mass-to-light ratio of the old disk , $\Upsilon
_*=(M/L)_{old\ disk}$, is approximately constant with the radius.
This means we'll take the projected surface density as
\begin{equation}
\sigma (r) =\Upsilon _* I(r) = \Upsilon _* I_0 e^{ - r/r_d } =
\sigma _0 e^{ - r/r_d },
\end{equation}
where $\sigma_0$ is the surface density at $r=0$.

Therefore, equation (6) could be reduced by substituting $\sigma
(r)$ with equation (12) to
\begin{equation}
s(\beta ) = \int_0^\infty  {rJ_0 (\beta r)\sigma (r)dr}  =
\Upsilon _* I _0 \frac{{r_d^2 }}{{(1 + r_d^2 \beta ^2 )^{3/2} }}.
\end{equation}

And the expression of $\langle {V_z^2}\rangle$ may be rewritten as
\begin{equation}
\begin{array}{c}
 \left\langle {V_z^2 } \right\rangle  = \frac{{\pi G}}{\alpha }\Upsilon _* I(r)[1 - P(r)] \\
\\
\ \ \ \ \ \ \ \ \ \ \ \ \  = \frac{{\pi G}}{\alpha }\Upsilon _* I_0 [e^{ - r/r_d }  - P_1 (r)] \\
 \end{array}
\end{equation}
where
\begin{equation}
\begin{array}{l}
 P_1 (r) = P(r)e^{ - r/r_d }  \\
 \\
\ \ \ \ \ \ \ \
  = \int_0^\infty  {\frac{{r_d^2 }}{{(1 + r_d^2 \beta ^2 )^{3/2} }}\beta [1 - \frac{1}{{(1 + \beta /\alpha )^2 }}]J_0 (\beta r)} d\beta  \\
 \end{array}
\end{equation}

Hence, if we know the $I_0$, $r_d$ from photometric study and
$\langle V_z^2 \rangle$ from spectroscopic observation, we can
calculate the thickness parameter $\alpha$ through equation (14)
and therefore obtain the thickness of galaxies.

\section{Application}
We found that there have existed extensively spectroscopic study
on two nearly face-on galaxies NGC 1556 and NGC 5247. Therefore,
we choose these two galaxies as examples to illustrate how our
method works.  The general information of these two galaxies are
summarized in Table 1.

\subsection{Photometric decomposition}
\setcounter{table}{0}
\begin{table*}
\begin{center}
\caption{Parameters of NGC 1566 and NGC 5247.}
\begin{tabular}{lccccc}
\hline\hline
NGC & Inclination & Classification
&Distance$^a$ (Mpc) &$M_H$ (mag) & ref \\
\hline
1566 &  28$^\circ$ $\pm$ 5$^\circ$ & Sc(s)I, Seyfert 1 & 17.4 & -24.0 &(1), (2), (4) \\
5247 &  $20^\circ
\pm 3^\circ$  & Sc(s)I-II & 16.0 & -23.2 &(1), (3), (4)\\
\hline
\end{tabular}
\begin{list}{}{}
\item[$^{\mathrm{a}}$]Obtained based on $H_0$=75 km s$^{-1}$
Mpc$^{-1}$.
\end{list}
References { (1) Sandage \& Tammann 1981; (2) Bottema 1992; (3)
van der Kruit \& Freeman 1986; (4) Jarrett et al. 2003.}
\end{center}
\end{table*}

\setcounter{figure}{0}
\begin{figure*}
\centering \epsfig{file=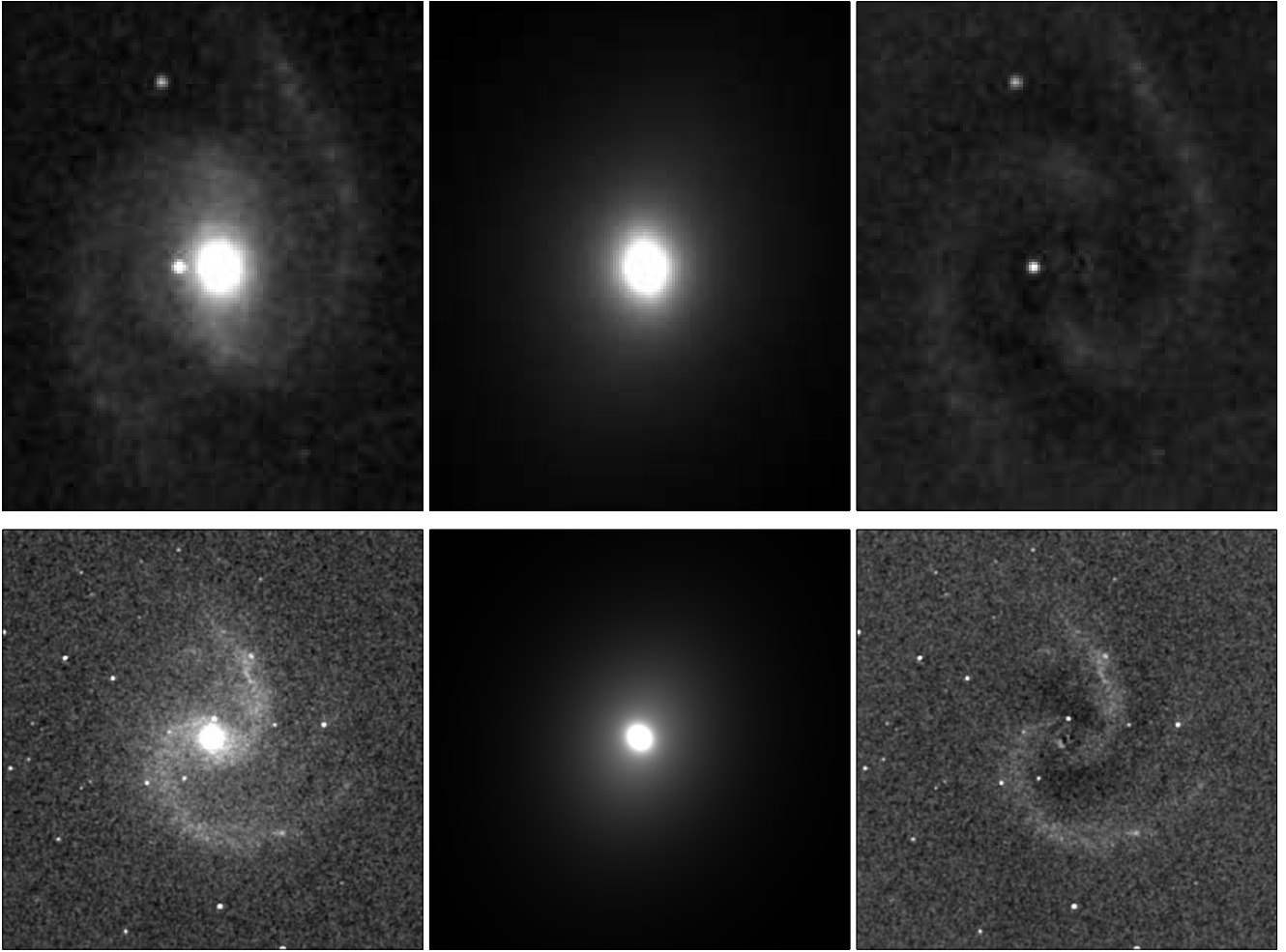} \caption{For both galaxies we
show the original images (\emph{left}), the smooth and symmetric
model images fitted by GALFIT (\emph{middle}) and the residual
images (\emph{right}). For NGC 1566 (\emph{top}), the size of each
field is $2'.4 \times2'.9$; for NGC 5247 (\emph{bottom}), the size
is $6'.7 \times 6'.7$} \label{all}
\end{figure*}

Although there also existed photometric results for NGC 1566 (de
Vaucouleurs 1973; Bottema 1992) and 5247 (van der Kruit \& Freeman
1986), we can not use these data directly since the photometric
decomposition is very important for our purpose. The photometric
decomposition can disentangle the contributions to the total
luminosity of the bulge and disk, and this will allow the
identification of the disk region. Moreover, because the infrared
images would suffer much less extinctions than optical ones do, we
use the 2MASS H-band images to do photometric measurements.

\begin{table*}
\begin{center}
\caption{Bulge-disk decompositions for NGC 1566 and NGC 5247. The
fitting functions are a Gaussian (for the nucleus), S\'{e}rsic
$r^{1/n}$ law (for the bulge) and an exponent (for the disk).}
\begin{tabular}{lccccccccc}
\hline\hline\noalign{\smallskip} \noalign{\smallskip}
&$M_{tot}$  & FWHM/$r_e$/$r_d$$^a$ & $\mu_e$/$\mu_0$$^b$&PA&&&&\\
Component &(mag)& (kpc)&(mag arcsec$^{-2}$)&($\circ$)&$n$&q&c&$\chi^2_\nu$\\
\hline\noalign{\smallskip} \multicolumn{9}{c}{NGC 1566}\\
\hline\noalign{\smallskip}
Nucleus& 11.70$\pm$0.03 &0.32$\pm$0.01&...& 88.02$\pm$2.36&...&0.89$\pm$0.01&0.26$\pm$0.11&\\
Bulge&9.25$\pm$0.02&0.92$\pm$0.03&17.05$\pm$0.06&12.21$\pm$0.48&1.65$\pm$0.03&0.79$\pm$0.00&-0.28$\pm$0.01&\\
Disk&7.65$\pm$0.00&3.29$\pm$0.04&17.20$\pm$0.04&-15.62$\pm$0.32&...&0.73$\pm$0.00& -0.29$\pm$0.02&0.310\\
\hline\noalign{\smallskip} \multicolumn{9}{c}{NGC 5247}\\
\hline\noalign{\smallskip}
Bulge& 10.69$\pm$0.01&0.67$\pm$0.02&18.06$\pm$0.04&42.20$\pm$1.44&1.24$\pm$0.01&0.87$\pm$0.00&0.01$\pm$0.04&\\
Disk&7.84$\pm$0.00&4.78$\pm$0.03&18.60$\pm$0.03&-18.72$\pm$0.70&...&0.80$\pm$0.00&-0.08$\pm$0.02&0.130\\
\hline
\end{tabular}
\end{center}
\begin{list}{}{}
\item[$^{\mathrm{a}}$] For the nucleus it is FWHM, for the bulge
it is $r_e$, and for the disk it is $r_d$; \item[$^{\mathrm{b}}$]
For the bulge it is $\mu_e$, and for the disk it is $\mu_0$.
\end{list}
\end{table*}

\begin{figure*}
\centering \epsfig{file=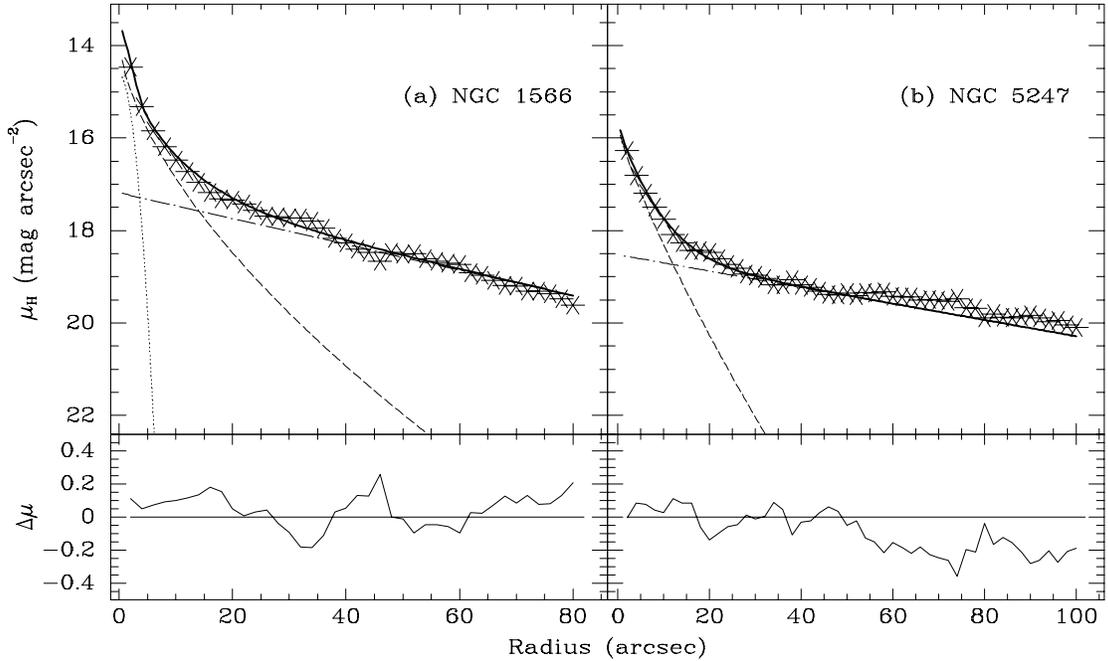} \caption{2MASS H-band radial
surface brightness profiles. Symbols show the radial profiles
obtained from ELLIPSE, the dotted line shows the Seyfert core
component fitted to the Gaussian function, the dashed lines show
the bulge components fitted to the S\'{e}rsic $r^{1/n}$ law, and
the dashed-dotted lines show the disk components fitted to the
exponential law. The thick solid lines are the superposition of
all the components. The lower portion of each panel shows the
difference between the data (ELLIPSE) and the fits (GALFIT).}
\label{all}
\end{figure*}

To obtain photometric parameters, $r_d$ and $\mu_0$, we use the
two dimensional image-fitting algorithm GALFIT (Peng et al. 2002)
designed to extract structural parameters directly from the galaxy
images. GALFIT assumes a two-dimensional model profile for the
galaxy. The functional forms of the models we choose to fit
include combinations of a Gaussian, a S\'{e}rsic $r^{1/n}$ law and
an exponential disk profile. For the case of a Seyfert 1 galaxy,
NGC 1566, we use all three functions to fit, while we use the last
two functions for NGC 5247. We fit the following: the (x, y)
position of the center, $M_{tot}$ (the total magnitude of the
component), $r_e$ (the effective radius), $n$ (the S\'{e}rsic
index), $q$ (the axis ratio), $r_d$ (the scale length of the
exponential disk), the major position angle (PA), and c (the
diskiness/boxiness index). The detailed parameters of all fitted
components are listed in Table 2. The original images, the smooth
and symmetric model images fitted by GALFIT and the residual
images are shown in the left, middle and right panels in Figure 1,
respectively.

In order to check whether or not the fitted results are reliable,
we examine the residual images obtained by subtracting the fitted
models. In both cases we found no obvious evidence (except for the
expectant spiral arms) for components beyond the fitted models.
Furthermore, we also measure surface brightness profiles using the
$IRAF\footnote{IRAF is distributed by the National Optical
Astronomy Observatories, which is operated by Associated of
Universities for Research in Astronomy, Inc., under cooperative
agreement with the National Science Foundation.}/ELLIPSE$
algorithm. In Figure 2 we show the radial profiles obtained from
the isophote fitting by $ELLIPSE$ and the profiles of the chosen
model for each galaxy. The differences between the observational
data ($ELLIPSE$) and the models (GALFIT) range mostly over
$\pm$0.2 mag arcsec$^{-2}$, except for the regions distorted by
the spiral arms at large radii.  So we conclude that our global
fits are reliable.

\subsection{Fitting the observed dispersion data}

The vertical velocity dispersions of NGC 1566 were gotten from
Table 4, column 4 in Bottema (1992); NGC 5247 from Table 1, column
4 in van der Kruit \& Freeman (1986). As these two galaxies are
not exactly face-on (see Table 1), this will affect the
observational data at two aspects: (1) the observed velocity
dispersion is not exactly equal to the vertical velocity
dispersion. The observed dispersion is approximately 4\% higher
than the real vertical dispersion for NGC 1566 (Bottema 1992),
$\sim$5\% larger for NGC 5247 (van der Kruit \& Freeman 1986). (2)
the observed surface brightness only amounts to about $cos i$
times the face-on brightness. Before using the photometric and
spectroscopic data for fitting, we have corrected the affections
of the inclination. We'll present a further discussion about how
the inclination affects the fitted results in the following.

To obtain the thickness parameter $\alpha$, we use a least-squares
method to fit the observational data shown in Figure 3, and 4, for
NGC 1566 and NGC 5247, respectively, and only the points with
radius larger than $\sim 2-3 r_e$, where the bulge's contribution
to the velocity dispersion may be neglect according to Terzi\'{c}
\& Graham (2005) (see Figure 11 and 12 in their paper). In the
process of fitting, we first adopt $\Upsilon_{*,H}$ (the
mass-to-light ratio for H-band) as an empirical value, following
Devereux et al. (1987), Oliva et al. (1995) and Giovanardi \& Hunt
(1996), and then use equation (14) to derive the thickness
parameter $\alpha$ and $\chi^2_\nu$. Whereafter, we adopt this
fitted $\alpha$ to derive the mass-to-light ratio
$\Upsilon_{*,H}$. Such an iteration are repeated until $\Delta
\chi^2_\nu / \chi^2_\nu$ ($\Delta \chi^2_\nu$ is the difference
between the $\chi^2_\nu$ given by two successive fitting) is less
than several times $10^{-4}$. The parameters needed in our fitting
process and the fitted results are listed in Table 3. The best
fitted results are also shown in Figure 3 and 4 with dotted lines.

\begin{table*}
\centering \caption{Photometric parameters and least-square fitted
results for NGC 1566 and NGC 5247.}
\begin{tabular}{lccccccc}
\hline\hline\noalign{\smallskip}
NGC&$I_0$&$\Upsilon_*$&&$\alpha$&$H$$^a$&Band\\
 &($L_\odot \ pc^{-2})$&($M_\odot/L_\odot$)&$\chi^2_\nu$&(kpc$^{-1}$)&(kpc)&\\
\hline\noalign{\smallskip}
 1566 & 1193.6$\pm$29.7&0.49$\pm$0.03    &0.48&1.75$\pm$0.56 &1.14$\pm$0.36   & H \\
 5247 & 328.7$\pm$4.5&0.91$\pm$0.03     &0.22&1.32$\pm$0.52 &1.52$\pm$0.60    & H \\
 \hline
\end{tabular}
\begin{list}{}{}
\item[$^{\mathrm{a}}$]Calculated according to $H=2/\alpha$.
\end{list}
\end{table*}

\begin{figure*}
\centering \epsfig{file=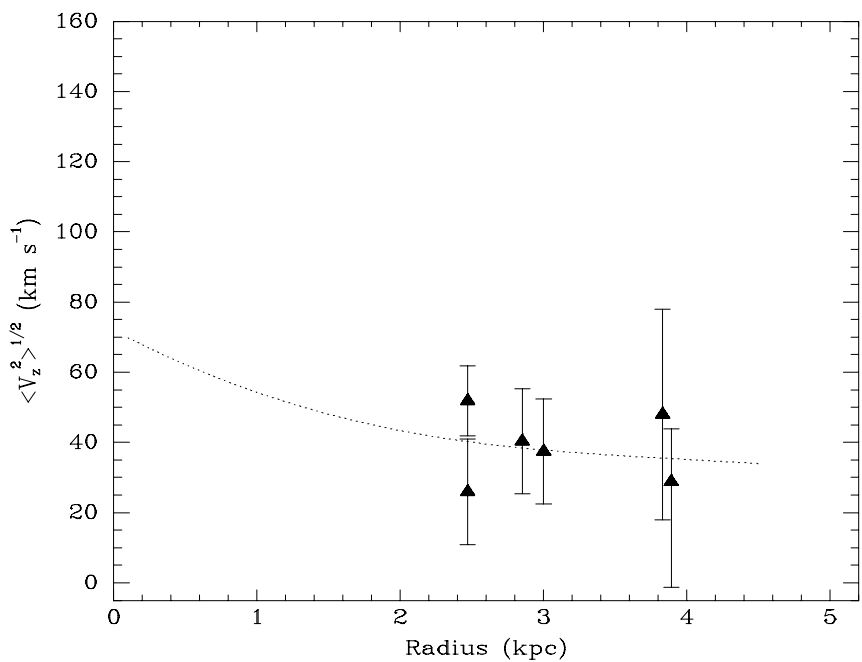} \caption{Observed (filled
triangles) and fitted (dotted line) distribution of vertical
stellar velocity dispersion of the disk in NGC 1566. The observed
data are from Bottema (1992), but the data which belong to the
Seyfert core and the bulge are not included in the figure, and not
taken into account for the determination of the thickness
parameter $\alpha$. The line is our least-squares fitted results
with equation (14) in the text.} \label{all}
\end{figure*}

\begin{figure*}
\centering \epsfig{file=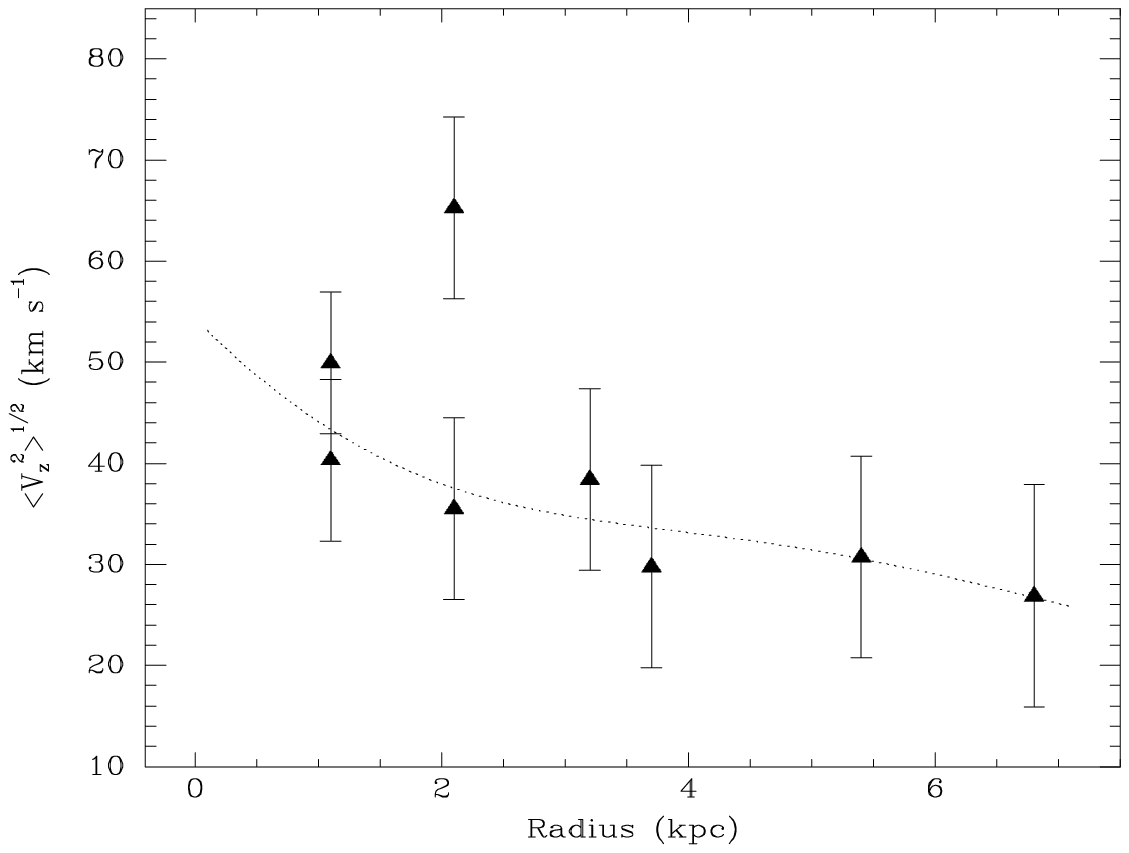}\caption{Observed (filled
triangles) distribution and our fitted results (dotted line) with
H-band photometric data for the disk of NGC 5247. The observations
are obtained by van der Kruit \& Freeman (1986). But the data
belonging to the bulge are excluded from this figure and not taken
into account for our fit.} \label{all}
\end{figure*}

The values of $\chi^2_\nu$ given by our least-squares fitting for
NGC 1566 and NGC 5247 are 0.48 and 0.22, respectively (see Table
3). These values of $\chi^2_\nu$ are somewhat smaller than
statistically acceptable, presumably because of the large
measurement errors of the observed data and the small samples used
for the fittings. However, the best-fitted results with our model
are very exciting, as seen from Fig 3 and 4. They show that our
method may be realistic and could be used universally to measure
the thicknesses of non-edge-on disk galaxies. What's more, our
method can derive the mass-to-light ratio simultaneously. But the
value gotten by this way are overestimated since the mass
contribution of the dark matter halo is neglected. However, this
is a reasonable assumption. The kinematic data cover a radial
range where luminous matter can be assumed to dominate the galaxy
dynamics. As shown in Table 3, the derived mass-to-light ratio for
these two galaxies are well consistent with the empirical values
(Devereux et al. 1987; Oliva et al. 1995; Giovanardi \& Hunt
1996). This also suggests that our method is reasonable.

When this method is applied to other measurements, we maybe adopt
an easier way that we only need to know the vertical velocity
dispersion at some exact radius (must be larger than $\sim 2r_e$),
$\langle V_{z,r_1}^2 \rangle^{1/2}$, and the photometric results,
e.g. the scale length of the disk $r_d$ and the central
extrapolated surface brightness $I_0$. Then, equation (14) can be
rewritten as
\begin{equation}
\ \left\langle {V_{z, r_1}^2 } \right\rangle \alpha  + \pi G
\Upsilon _* I_0 P_1(r_1, \alpha)- \pi G \Upsilon _* I_0 e^{ -
r_1/r_d } = 0
\end{equation}
The thickness parameter, $\alpha$, is the only unknown parameter
in this equation if we have assumed the mass-to-light ratio.
Alternatively, we can use the surface density-weighted average
velocity dispersion, $\overline{\left\langle {V_{z}^2 }
\right\rangle}$, to estimate the thickness parameter $\alpha$
through the following equation,
\begin{equation}
\overline {\left\langle {V_z ^2 } \right\rangle }  =
\frac{{\int_0^{r_d } {\frac{{\pi G}}{\alpha }\sigma^2 (r)[1 -
P(r)] r dr} }}{{\int_0^{r_d } {\sigma (r) r dr} }}.
\end{equation}
When equation (17) is used for measurements, equation (12) would
be substituted for $\sigma (r)$.

\section{Discussion}

\begin{table*}
\centering \caption{Parameters of the near face-on galaxies from
our study compare with those of edge-on galaxies from the study by
van der Kruit \& Searle.}
\begin{tabular}{lcccccc}
\hline\hline
NGC &RC3 Type &$M_H$$^a$ &$r_d$$^b$ (kpc) &$H$$^c$ (kpc)& $r_d/H$ &ref \\
\hline
 1566 &Sbc & -24.0 & 3.3  &1.1      &3.0  &our study\\
 5247 &Sbc & -23.2 & 4.8  &1.5      &3.2  &our study\\
 891  &Sb  & -23.3 & 4.9  &0.99     &5.0     &(1)\\
 4013 &Sb  & -22.4 & 3.4  &1.1      &3.1     &(2)\\
 4217 &Sb  & -22.9 & 3.5  &1.7      &2.1     &(2)\\
 4244 &Scd & -20.7 & 2.6  &0.58     &4.5     &(3)\\
 4565 &Sb  & -23.7 & 5.5  &0.79     &7.0     &(3)\\
 5023 &Scd & -19.5 & 2.0  &0.46     &4.3     &(2)\\
 5907 &Sc  & -23.1 & 5.7  &0.83     &6.9     &(2)\\
 7814 &Sab & -23.5 & 8.4  &2.0      &4.2     &(4)\\
\hline
\end{tabular}
\begin{list}{}{}
\item[$^{\mathrm{a}}$] The magnitudes are all gotten from Jarrett
et al. 2003; \item[$^{\mathrm{b}}$]$r_d$ in KS is represented by
$h$; \item[$^{\mathrm{c}}$]$H$ is represented by $z_0$, which is
obtained by fitting the light distribution with
$\mu(R,z)=~\mu(0,0)(R/h) K_1(R/h)$sech$^2(z/z_0)$, $K_1$ is the
modified Bessel function, in KS.
\end{list}
References { (1) van der Kruit \& Searle 1981b; (2) van der Kruit
\& Searle 1982a; (3) van der Kruit \& Searle 1981a; (4)van der
Kruit \& Searle 1982b.}
\end{table*}

\begin{figure*}
\centering \epsfig{file=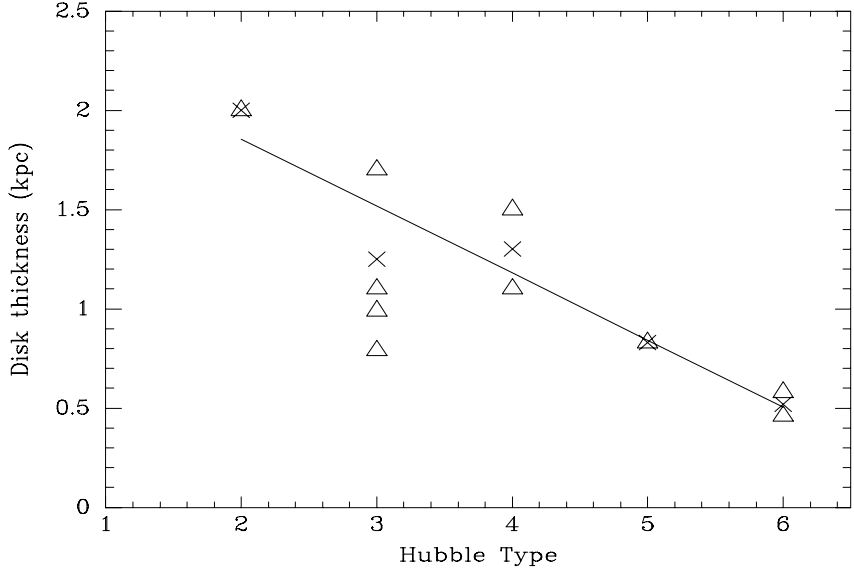}\caption{Thickness plotted versus
Hubble type(1, 2, 3, 4, 5 and 6 for Sa, Sab, Sb, Sbc, Sc and Scd,
respectively.) The continuous line is the fitted result of the
median value of each type (represented by the crosses).  }
\label{all}
\end{figure*}

So far, we have proposed an method to determine the thickness of
the non-edge on disk galaxies $H$, based on the photometric
result, scale length of disk ($r_d$) and extrapolated central
surface brightness ($I_0$); and spectroscopic result, the vertical
stellar velocity dispersion ($\left\langle {V_z^2 }
\right\rangle$). The fundamental assumption required to derive the
conclusions presented here is that the disk mass profile is
exponential, both along the vertical direction (i.e.,
$z$-direction, see equation (3)), which is different from the
isothermal model used by KS, and along the radial direction (i.e.,
$r$-direction, see equation (1)). It has been shown by de Grijs \&
van der Kruit (1996) that the best fitting vertical model is more
peaked than expected for an isothermal sheet distribution and the
simple exponential fits turned out to be good approximations of
the vertical surface brightness profile. Peletier \& de Grijs
(1997) also find that the $K$-band light (often assumed to trace
mass) indeed has a very peaky, almost exponential, vertical
distribution in edge-on spirals. By contrast, Gould, Bahcall \&
Flynn (1996) find that the vertical distribution of Galactic M
dwarfs (which probably trace the stellar disk) is somewhat less
peaky.

Because our method is intended to work for a pure disk galaxy, the
photometric decomposition is another important ingredient
affecting the fitted result. As described in Section 3.1,
photometric decomposition allows us to identify the real disk
region and the range that the bulge/nucleus (if the galaxy has)
affects. We also use the data \emph{just} outside the effective
radius $r_e$ to fit $\alpha$ for NGC 1566, and we obtain a much
bigger $\chi^2$ of 1.8 and a smaller $\alpha$ of 0.96 kpc$^{-1}$.
For NGC 5247, we use the data out to the center since there are no
data just near the effective radius. This also gives a bigger
$\chi^2$ of 0.35 and a smaller $\alpha$ of 1.0 kpc$^{-1}$. This
suggests that we need to select the radial range according to the
total mass of the bulge/nucleus. The inclination is another
ingredient affecting the fitting results, particularly true for
the mass-to-light ratio. Applied corrections for different
inclinations will result in different mass-to-light ratio, but
almost the same thickness parameter. For the same velocity
dispersion data, a smaller inclination gives a larger
mass-to-light ratio.

The parameter $H$ should be regarded as about twice the scale
height of the best fit of an exponential to the actual
distribution. We've compared our some relevant results, e.g.
$r_d/H$ (the scale length to the effective height ratio), $H$ (the
effective height) of these three galaxies with those of eight
edge-on spirals of KS, which are directly measured by fitting the
photometric data. As shown in Table 4, our results are very
consistent with KS. This make us believing that our new method for
determining the thickness of non-edge-on disk galaxies is
reasonable and feasible. In Figure 5 we plot the thickness versus
the Hubble type. The continuous line is the fitted result for the
crosses, which represent the median value of the thickness for
each type. The trend that the thickness of disks decreases along
the Hubble sequence is also found by Ma (2002) and Zhao, Peng \&
Wang (2004). However, it needs further study on this correlation
because the sample used here is rather small.

The method presented in this paper is rather difficult to be used
for measurements of large galactic samples because the
distributions of the vertical stellar velocity dispersions are
inconvenient to be obtained in practice. Fortunately, the most
essential purpose of the present paper is to test the availability
of the method proposed by Zhao et al. (2004), which is on the
basis of the density wave theory. Hence, we measured the thickness
of the disks of NGC 1566 and NGC 5247 used the method put forward
by Zhao et al (2004) and obtained $0.6\pm0.4$ kpc and $0.7\pm0.5$
kpc, for NGC 1566 and NGC 5247, respectively. These results are
somewhat smaller than the results obtained by using the method
shown in this paper. However, these values are also in the range
of the measurement error. Therefore, the method presented in Zhao,
Peng \& Wang (2004), which is more easily to be carried out, might
be available and might be used to measure the thickness of the
disks of nearly face-on galaxies for large samples.

\begin{acknowledgements}
The authors are very grateful to Dr. E. M. Corsini for his
thoughtful and instructive comments which significantly improved
the content of the paper. We are grateful to Dr. Xinlian Luo and
Dr. Qiusheng Gu of the Department of Astronomy, Nanjing University
for valuable discussions and suggestions. We also thank Dr. T. H.
Jarrett for his kindly help. We would like to express our
appreciation for the support by the Chinese National Science
Foundation No. 10573011, 10273006 and the Doctoral Program
Foundation of State Education Commission of China. This paper uses
data products from 2MASS, a joint project of the University of
Massachusetts and the IPAC/CalTech, funded by NASA and NSF.
\end{acknowledgements}

\appendix

\section{Mathematical basis}
In order to study the effect of finite thickness of the disk on
the dynamical properties of disk galaxies, it is convenient to
adopt cylindrical coordinates $(r,\theta,z)$  with the galactic
center chosen at the origin of the coordinate system. The galactic
plane is depicted by $z=0$ and $\theta$  represents the azimuthal
coordinate. For disk galaxies with zero-thickness, the
self-gravitational potential $\Phi$ is governed by Poisson's
equation
\begin{equation}
\nabla ^2 \Phi  = 4\pi G\sigma (r,\theta )\delta (z),
\end{equation}

where $G$ is the gravitational constant, $\nabla^{2}$ denotes the
Laplacian operator, $\delta(z)$ is Dirac's delta function that
confines all the disk mass to the galactic plane and
$\sigma(r,\theta)$ represents the surface density. In regions
outside the galactic plane $z\neq0$, equation (A.1) then becomes
\begin{equation}
\nabla ^2 \Phi (r,\theta ,z) = 0.
\end{equation}
since there is no distribution of matter outside the disk plane.
It is particularly convenient at this point to introduce the
Laplace transform for the vertical coordinate $z$  and the Fourier
transform for the azimuthal angle $\theta$, namely
\begin{equation}
\Phi (r,\theta ,z) = e^{ - im\theta } \int_0^\infty  {U_\beta
(r)e^{ - \beta \left| z \right|} d\beta }.
\end{equation}
Applying these transformations to equation (A.2) we then have
\begin{equation}
x\frac{d}{{dx}}(x\frac{{dU_\beta  (x)}}{{dx}}) + (x^2  - m^2
)U_\beta  (x) = 0,
\end{equation}
where  $x=\beta r$, and the solution to equation (A.4) is the well
known Bessel function of order m
\begin{equation}
U_\beta  (x) \equiv U(\beta r) =  - J_m (\beta r).
\end{equation}
Using equation (A.5) in equation (A.3), we obtain
\begin{equation}
\Phi (r,\theta ,z) = e^{ - im\theta } \int_0^\infty  {\left[ { -
J_m (\beta r)} \right]e^{ - \beta \left| z \right|} d\beta }.
\end{equation}
On the other hand, we may integrate Poisson's equation (equation
(A.1)) with respect to $z$ to obtain the following expression for
the surface density $\sigma(r,\theta)$
\begin{equation}
\sigma (r,\theta ) = \frac{1}{{4\pi G}}\{ [\frac{{\partial \Phi
}}{{\partial z}}]_{0^ +  }  - [\frac{{\partial \Phi }}{{\partial
z}}]_{0^ -  } \}.
\end{equation}
Substituting equation (A.6) in equation (A.1) and equation (A.7)
we can recast Poisson equation for disk galaxies with zero
thickness in the form
\begin{equation}
\begin{array}{l}
\nabla ^2 \left\{ {\int_0^\infty  {e^{ - im\theta } \left[ { - J_m
(\beta r)} \right]e^{ - \beta \left| z \right|} d\beta } }
\right\} \\
\\
\ \ \ = \int_0^\infty  {e^{ - im\theta } 2\beta J_m (\beta
r)\delta (z)d\beta },
\end{array}
\end{equation}
\begin{equation}
\nabla ^2 [J_m (\beta r)e^{ - im\theta  - \beta \left| {z - z'}
\right|} ] =  - 2\beta J_m (\beta r)e^{ - im\theta } \delta (z -
z').
\end{equation}

\section{Solution of Poisson's equation for an axisymmetric disk}
On the basis of the mathematical treatment just depicted together
with the standard method of Green functions, it is straightforward
to derive the appropriate Poisson equation for the
three-dimensional disk galaxies with finite thickness
\begin{equation}
\nabla ^2 \varphi (r,z) = 4\pi G\rho (r, z)
\end{equation}
where the vertical distribution of matter is depicted by
Parenako's law, $\rho (r, z) = \frac{\alpha }{2}\sigma (r)e^{ -
\alpha \left| z \right|}$ (see equation (3) in the text). To
proceed, we apply the Bessel-Fourier transform to $\sigma (r)$ to
obtain
\begin{equation}
\sigma (r) = \int_0^\infty  {\beta J_0 (\beta r)S (\beta )d\beta
},
\end{equation}
\begin{equation}
S (\beta ) = \int_0^\infty  {rJ_0 (\beta r)\sigma (r)dr}.
\end{equation}
where $S(\beta )$ is the Bessel-Fourier transform for $\sigma
(r)$. Substituting equations (A.1) and (B.2) into equation (B.1),
the three-dimensional Poisson equation can be rewritten as
\begin{equation}
\begin{array}{l}
\nabla ^2 \varphi (r, z) = 2\pi G\alpha\\
\\
\ \ \ \ \ \ \ \ \ \ \ \ \ \ \  \cdot \int_{ - \infty }^\infty {e^{
- \alpha \left| {z'} \right|} dz'\int_0^\infty  {\beta J_0 (r\beta
)S(\beta )\delta (z - z')d\beta } }.
\end{array}
\end{equation}
Compare equation (B.4) with equation (A.9) we then find the formal
solution of the gravitational potential for a galactic disk with
scale height $H=2/\alpha$
\begin{equation}
\varphi (r, z) =  - 2\pi G \alpha \int_0^\infty {J_0 (r\beta )S
(\beta )F(\alpha ,\beta ,z)d\beta },
\end{equation}
where
\begin{equation}
F(\alpha ,\beta ,z) = \frac{1}{{\beta ^2  - \alpha^2}}\left[
{\beta e^{ - \alpha \left| z \right|}  - \alpha e^{ - \beta \left|
z \right|} } \right].
\end{equation}

\end{document}